\documentstyle[epsfig,12pt]{article}

\def\Journal#1#2#3#4{{#1} {\bf #2}, #3 (#4)}

\def\NPA{{Nucl. Phys.} A}
\def\NPB{{Nucl. Phys.} B}

\def\PRL{Phys. Rev. Lett.}
\def\PRC{{Phys. Rev.} C}
\def\PRD{{Phys. Rev.} D}

\newfam\BMath
\font\BMathL=cmmib10 
\font\BMathl=cmmib7
\font\BMathm=cmmib5
\textfont\BMath=\BMathL \scriptfont\BMath=\BMathl
\scriptscriptfont\BMath=\BMathm

\def\a{\alpha}
\def\b{\beta}

\def\e{\epsilon}

\def\p{\pi}
\def\q{\theta}

\def\be{\begin{equation}}
\def\ee{\end{equation}}
\def\bea{\begin{eqnarray}}
\def\eea{\end{eqnarray}}

\def\fref#1{Fig.~\ref{#1}}
\def\bfi{\begin{figure}}
\def\efi{\end{figure}}

\newcommand{\ncom}{\newcommand}
\ncom{\vo}[1]{{\fam\BMath #1}}
\ncom{\vt}[2]{({\fam\BMath #1}+{\fam\BMath #2})}
\ncom{\vmo}[1]{\vert{\fam\BMath #1}\vert}
\ncom{\vmt}[2]{\vert\mbox{\bf #1}+\mbox{\bf #2}\vert}
\ncom{\lan}{\langle}
\ncom{\ran}{\rangle}
\ncom\nonum{\nonumber \\}
\ncom\fx{\!\!\!\!}
\ncom\gsim{\mbox{\raisebox{-0.6ex}{\ $\stackrel {>}{\sim}$\ }}}
\ncom\lsim{\mbox{\raisebox{-0.6ex}{\ $\stackrel {<}{\sim}$\ }}}

\ncom{\half}{{1\over 2}}
\ncom{\third}{{1\over 3}}
\ncom{\fourth}{{1\over 4}}
\ncom{\fifth}{{1\over 5}}
\ncom{\sixth}{{1\over 6}}

\ncom\Tg{T_{eq\; g}}
\ncom\Tq{T_{eq\; q}}
\ncom\qg{\q_g}
\ncom\qq{\q_q}

\value{textfraction}

\font\lbigbf=cmbx10 scaled 1200

\begin{document}

\begin{flushright}
\footnotesize \sffamily WU B 97/10
\end{flushright}

\begin{centering}
{\lbigbf THE EQUILIBRATION OF A PARTON PLASMA CREATED \\
IN RELATIVISTIC HEAVY ION COLLISIONS}

\vspace{0.25cm}

{S.M.H. Wong}

\em Fachbereich Physik, Universit\"at Wuppertal,
D-42097 Wuppertal, Germany

\end{centering}

\begin{abstract}

We study the equilibration of a parton plasma in
terms of its parton compositions and its state
of thermalization. In studying the evolution of the 
plasma, one has to assume a small value of the strong 
coupling constant. This value is by no means fixed. 
By varying this only parameter in our calculation, we 
show the dependence of equilibration on its magnitude. 
It is shown that both kinetic and parton equilibration are 
faster with increasing coupling but the plasma cools much 
more rapidly resulting in shortened lifetime. The degree
of equilibration improves significantly for quarks and 
antiquarks but not so for gluons and the total 
generated entropy is reduced.
With a coupling depending on the average parton 
energy, there is additional acceleration in the 
equilibration during the evolution.

\end{abstract}
\vspace{-0.5cm}

\section{Introduction}
\label{sec:intro}

Equilibration in relativistic heavy ion collisions is
an important problem because particle signatures upon
which one relies for detecting the deconfined matter are
directly influenced by the temporal development of the
remnant of the initial collisions. The evolution of
the so produced secondary partons will undergo
many interactions and hence in accordance
with the laws of thermodynamics, the plasma consisting 
of these partons will try to reach equilibrium. 
This process is not without hindrance
and there is no guarantee that equilibrium can be
reached. First because it is a highly compressed
system at the beginning, it will try to
push itself apart and therefore undergoes expansion,
which is very disruptive for the equilibrating parton
system. In order to equilibrate, the net interaction
rate must dominate over the expansion rate. Second,
time is limited because a sufficiently cooled system
will not be able to resist the confining force which
is also responsible for the equilibration process in the first
place. In this talk, we would like to point out that
the confining force actually helps the parton system
to equilibrate before proceeding to change the
very form of the components of the system through the
deconfinement phase transition and hence ending the 
pure partonic equilibration process all together. 

In previous studies of the equilibration, in chemical
\cite{biro&etal1,shury&xion} as well as in kinetic 
equilibration \cite{shury1,geig,wong1}, the system evolved through 
a period of time varying from several fm/c to over 10 fm/c
depending on the initial conditions. In this period,
the estimated temperature dropped by hundreds of MeV and the
average parton energy also decreased by over 1 GeV. So the
system underwent significant changes. In these studies,
a value of $\a_s=0.3$ was used which is equivalent to an
average momentum transfer of $Q \sim 2$ GeV and 
$\Lambda_{QCD} \sim 200$ MeV. But we have just pointed out
that the average parton energy varies by so much that we
cannot reasonably expect the average $Q$ to remain at
2 GeV. So during the evolution of the plasma, the strength
of the interaction is also likely to evolve with the system.
We shall try to take this into account. A second point
also related to the coupling in the question of the
equilibration of the parton plasma is the fact that due
to screening and the generation of medium masses in a 
dense environment \cite{klim&weld}, no arbitrary infrared 
cutoff is required in the calculation. Therefore $\a_s$
is the only parameter, apart from the obvious initial
conditions, that one can choose. We would like to find out
how the results will be affected by the choice of $\a_s$. 
Also, as we have just mentioned, if we relate the coupling 
to the average momentum transfer of the system 
then even $\a_s$ is determined by the
system and no other parameter other than the initial
conditions remain. 

We shall use different values of $\a_s$ and
in addition we also assume
$Q$ to be given by the average parton energy and use
the one-loop formula 
$\a_s(Q) =4\p/\b_0 \ln(Q^2/\Lambda^2_{QCD})$ to obtain a
time-dependent coupling which varies as the system evolves 
\cite{wong2}.
We shall refer to this coupling in the following as $\a_s^v$.
The results of this evolution are then compared with those
obtained with fixed $\a_s$.

\section{The Evolution of a parton plasma with different
couplings}

\bfi
\centerline{
\hbox{
\epsfig{file=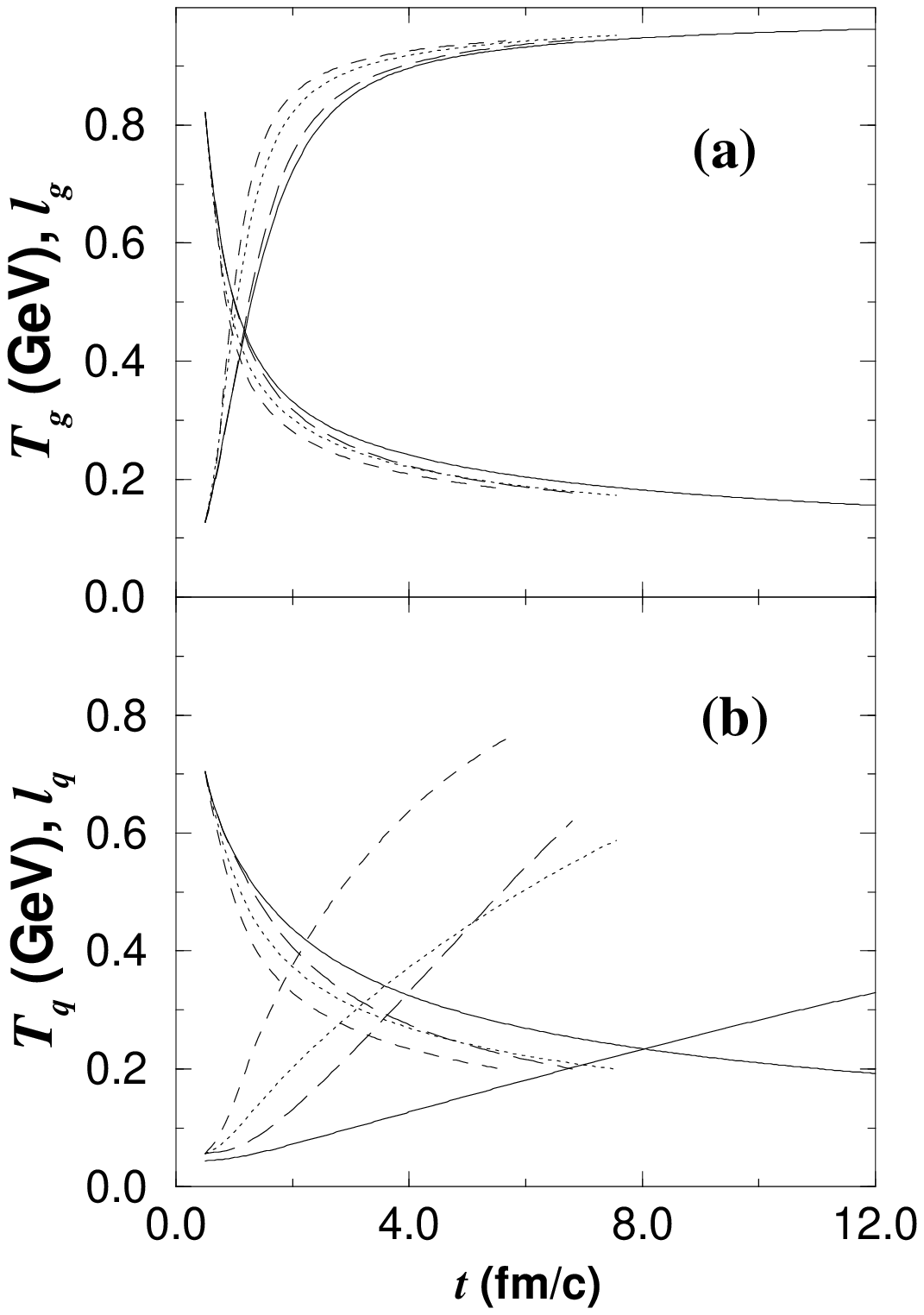,width=2.3in}
\epsfig{file=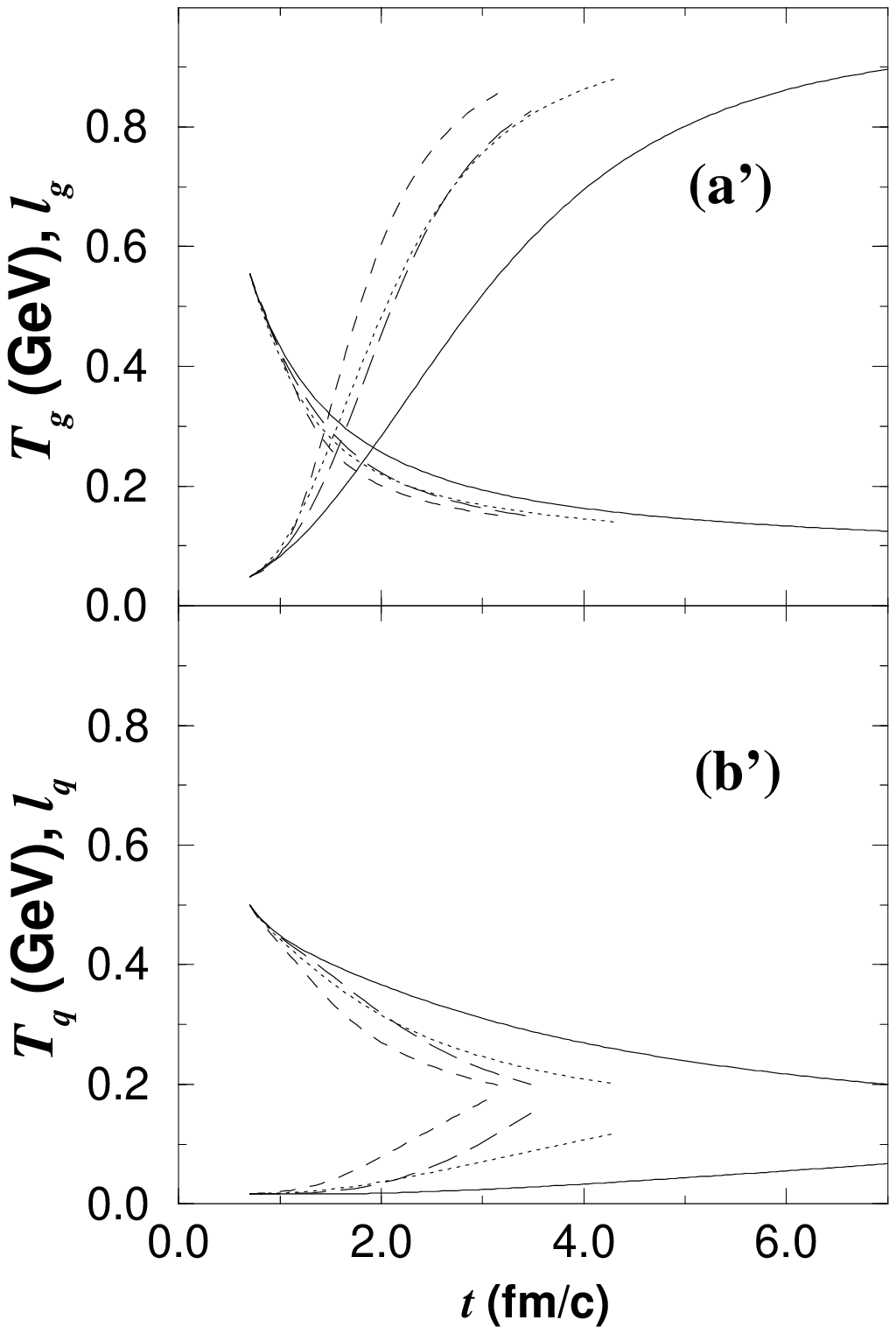,width=2.3in}
}}
\caption{The variations of the fugacities and the estimated 
temperatures with time and $\a_s$. Solid, dotted, dashed 
and long dashed lines are for $\a_s=0.3, 0.5, 0.8$ and 
$\a_s^v$ respectively. (a) for gluon and (b) for quark at LHC
and (a'),(b') are the same at RHIC.}
\label{fig:fugtem}
\efi

In order to determine the effects of the variation of
$\a_s$ on the equilibration. We choose
values of $\a_s=0.5,0.8$ and $\a_s^v$ in addition to
$\a_s=0.3$ \cite{wong1,wong2}. 
Such large values of $\a_s$ are chosen deliberately
to make the effects manifest. We are not after 
quantitative but rather qualitative results. In any case,
the uncertainties in the initial conditions do not
permit us to make any meaningful numerical predictions
at present. Such uncertainties will not concern us here,
we concentrate rather on the coupling. With
the same initial HIJING inputs \cite{gyu} as before, 
the evolution is performed using the method described 
in \cite{wong1}. 

To check the state of the equilibration, we look at the 
parton fugacities $l$, the longitudinal to transverse 
pressure ratios $p_L/p_T$ and the temperature estimates $T$. 
The first give us information about the partonic 
composition of the plasma, the second reveal the state 
of the kinetic equilibration of the system and the last 
tell us about the possible lifetime of the parton plasma. 
These are plotted in \fref{fig:fugtem} and \fref{fig:pres}.

Let us first look at the partonic composition of the plasma
as a function of time. These are shown in \fref{fig:fugtem}
(a),(b) for LHC and (a'),(b') for RHIC in terms of fugacities.
The curves shift towards the left upper corner with increasing
coupling. As can be seen, more prominant in the case of 
quark and antiquark than that of gluon, chemical equilibration 
is faster with increasing coupling. The final $l_g$ for gluon
are approximately the same, but for quark and antiquark,
larger $\a_s$ does make a difference in the final $l_q$.
A factor of 1.76 or more has been gained both at LHC and at 
RHIC. 

Turning to kinetic equilibration of the parton, we show
the degree of isotropy of the momentum distribution of each
parton component. This is done in terms of the ratio of the 
longitudinal to transverse pressure. 
\bfi
\centerline{
\hbox{
\epsfig{file=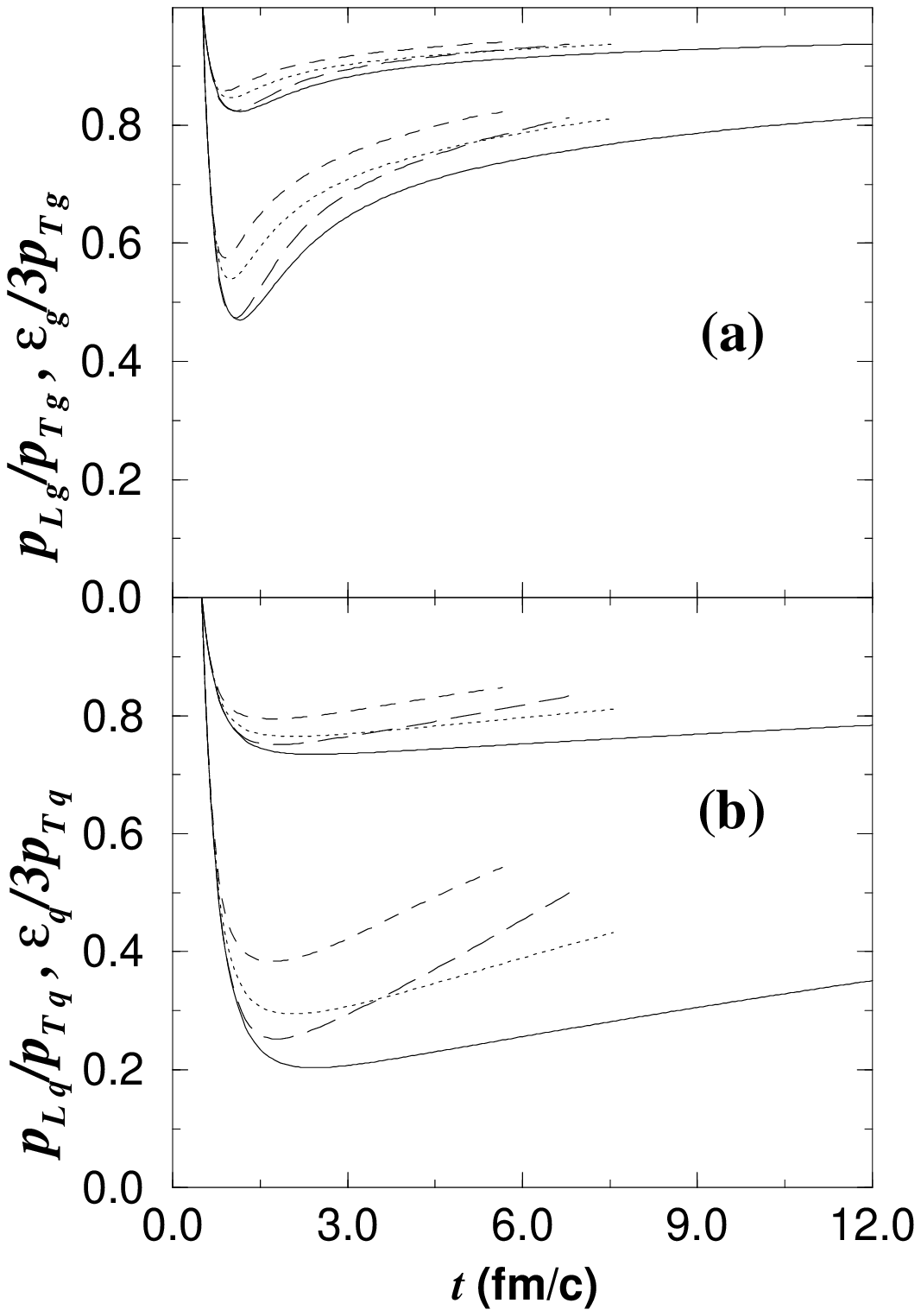,width=2.3in}
\epsfig{file=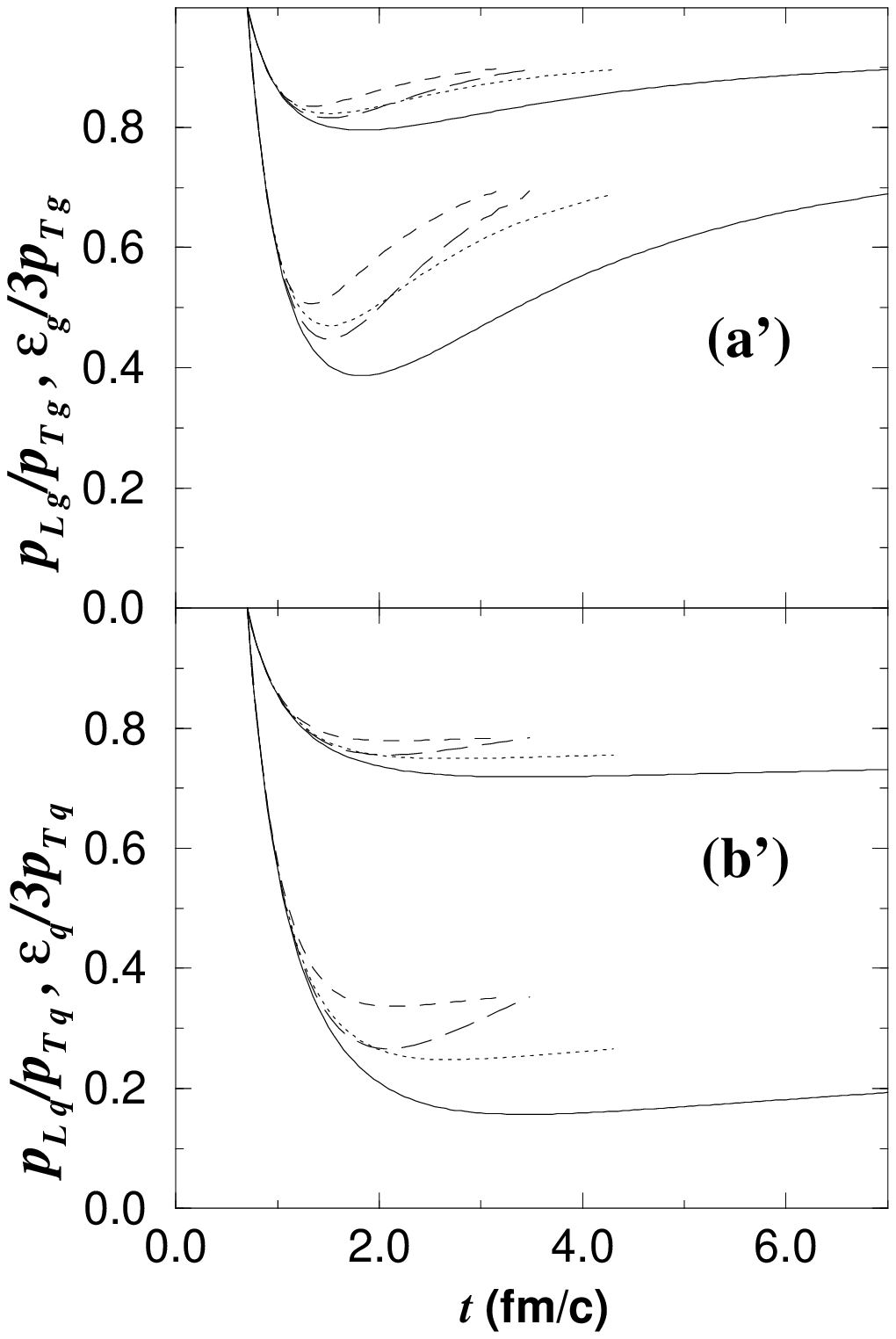,width=2.3in}
}}
\caption{The variations of the ratios of longitudinal and a third of 
the energy density to transverse pressure with 
time and $\a_s$. Solid, dotted, dashed and long dashed lines  
are for $\a_s=0.3, 0.5, 0.8$ and $\a_s^v$ respectively.}
\label{fig:pres}
\efi
These plots are shown
in \fref{fig:pres} (a),(b) for LHC and (a'),(b') for RHIC.
With increasing $\a_s$, the curves shift upwards,
that is closer to kinetic equilibrium or isotropic
momentum distribution. The final degree of kinetic 
equilibration for gluon is, like $l_g$, again approximately
the same. That for quark and antiquark is again much obvious
and shows improvement. The top set of curves in each plot is used 
as a double check which are plots of $\e/3 p_T$. These ratios
should all go to 1.0 at equilibrium. It now becomes obvious
that larger $\a_s$ speeds up equilibration for both
quark and gluon, but only quark and antiquark show signs of 
much obvious improvements in the degree of equilibration. All 
these are at the expense of shorter lifetime as one can see 
the curves with larger $\a_s$ are stopped at earlier times.
This is because the temperature estimates of the partons drop
faster with increasing $\a_s$. This is also plotted in 
\fref{fig:fugtem}. The effect of larger $\a_s$ is to reduce
the lifetime but speed up equilibration. Only quark and 
antiquark show significant improvements but not gluon. We 
are especially interested in the case of $\a_s^v$ 
because as stated in the introduction, the coupling 
should be affected by the evolution
of the plasma. As seen in the figures, the results start
off staying close to those of $\a_s=0.3$ but very soon
depart and shift across the constant $\a_s$ ``contours''.
So we see that with a time-dependent coupling as 
determined by the system, there is an acceleration effect.
As mentioned earlier, the confining force helps the
equilibration process but at a price of earlier
onset of the deconfinement phase transition.

Different strengths of the interactions also affect the
generated entropy. Larger $\a_s$ increases the gluon entropy
loss by gluon conversion into quark and antiquark pairs.
As we have seen the faster equilibration of quark and
antiquark, which leads also to a reduction in their 
generated entropy. The tendency is then
a reduced total entropy and hence a reduced final
pion multiplicity. If there is a first order phase
transition, the mixed phase will also be shortened. 

In summary, we have seen that larger coupling has much
more obvious effects on the degree of equilibration of
quarks and antiquarks than on gluons. Significant improvement
in the case of quarks and antiquarks but not in that of 
gluons. Both the lifetime and the total generated entropy
are sensitive to the strength of the interactions.
A time-dependent coupling based on the actual situation of the
system will lead to accelerated equilibration but the earlier 
arrival of the deconfinement phase transition.

\section*{Acknowledgements}

The author would like to thank the organizers for all 
their efforts towards this conference.

\end{document}